\documentclass[a4paper,12pt]{article}
\usepackage{float}
\usepackage{amsmath}
\usepackage{amssymb}
\usepackage{caption}
\usepackage{graphicx}
\usepackage{amsfonts,ulem}
\usepackage[colorlinks,linkcolor=blue,anchorcolor=red,citecolor=green]{hyperref}
\usepackage{cite}
\usepackage{geometry}
\usepackage{color}

\numberwithin{equation}{section}
\begin{document}

\allowdisplaybreaks
\normalem

\title{ Deforming  black holes with odd multipolar differential rotation boundary }

\vskip1cm

\author{Shuo Sun\footnote{sunsh17@lzu.edu.cn},  Tong-Tong Hu\footnote{hutt17@lzu.edu.cn},
 Hong-Bo Li\footnote{lihb2017@lzu.edu.cn},
 and  Yong-Qiang Wang\footnote{ yqwang@lzu.edu.cn, corresponding author}
\\ \\
   Research Center of Gravitation $\&$
 Institute of Theoretical Physics $\&$ \\
Key Laboratory for Magnetism and Magnetic of the Ministry of Education,\\ Lanzhou University, Lanzhou 730000, China
 \\
}

\maketitle

\begin{abstract}
Motivated by the novel asymptotically  global AdS$_4$ solutions with deforming horizon in [JHEP {\bf 1802}, 060 (2018)], we analyze
the boundary metric  with odd multipolar differential rotation  and numerically construct a family of deforming solutions with tripolar differential rotation boundary,
including two classes of solutions: solitons and black holes.
We find that the maximal values of the rotation parameter $\varepsilon$, below which the stable large black hole solutions could exist, are not a constant for $T> T_{schw}=\sqrt{3}/2\pi\simeq0.2757$. When temperature is much higher than $ T_{schw}$, even
though the norm of Killing vector $\partial_{t}$ keeps timelike for some regions of $\varepsilon<2$, solitons and black holes with tripolar differential rotation could be unstable and   develop hair due to superradiance.
As the temperature $T$ drops toward  $T_{schw}$,
 we find that though there exists  the  spacelike Killing vector $\partial_{t}$  for some regions of $\varepsilon>2$,  solitons and black holes still exist and do not develop hair due to superradiance.
 Moreover, for $T\leqslant T_{schw}$, the curves of entropy firstly combine into one curve and then separate into two curves again, in the case of each curve there are  two solutions at a fixed value of $\varepsilon$. In addition, we study the deformations of horizon for black holes by using an isometric embedding  in the hyperbolic three-dimensional space.
Furthermore,  we also study the quasinormal modes of  the solitons and black holes,
which have analogous behaviours to that
of dipolar rotation and quadrupolar rotation.
\end{abstract}

\newpage

\tableofcontents

\section{Introduction}
\hspace*{0.6cm}The uniqueness theorem of black holes \cite{a1,a2,a3,a4} in classical general relativity has shown that the asymptotically flat black hole solutions with zero angular momentum in four dimensions are named as Schwarzschild black holes, which have a spherical event horizon. It is well known that in four-dimensional anti-de Sitter (AdS) spacetime, there exist some solutions with  noncompact planar, hyperbolic horizons and compact horizons of arbitrary genus $g$. Recently, the asymptotically AdS black holes in the context of anti-de Sitter/comformal field theory (AdS/CFT) correspondence\cite{a5,a6,a7,a8} have aroused great interest, and it is especially important to study the physical properties and applications of such black holes.\\
\hspace*{0.6cm}Considering a asymptotically AdS black hole has a conformal boundary at infinity, the boundary metric of black hole could deform and thus obtain a black hole with deforming horizon, it is to mean that the curvature of the horizon is not a constant and the horizon could have various deformations. There are two methods to get the solutions of deforming black holes, one method is analytically constructed a family of the hyperbolic AdS black holes\cite{a9} with deforming horizon in four-dimensional spacetime by using the AdS C-metric\cite{a10,a11,a12}.  In addition, a class of four-dimensional AdS black holes with noncompact event horizons of finite area is found in
\cite{Gnecchi:2013mja,Klemm:2014rda}, and black holes with bottle-shaped horizons are found in\cite{a13}. The another method is to study the deforming black holes by using numerical methods. In \cite{r1}, the author constructed a family of deforming solutions with differential rotation boundary, including the solitons and black holes. These solutions have nontrivial boundary metrics with a dipoalr differential rotation profile
\begin{equation}
\Omega(\theta)=\varepsilon \cos(\theta),
\end{equation}
where the parameter $\varepsilon$ is the boundary rotation parameter and the polar angle $\theta$ is restricted to the interval $(0,\pi)$. The rotation profile $\Omega(\theta)$ is anti-symmetric on the equatorial plane $\theta=\pi/2$. When the rotation boundary parameter $\varepsilon$ is larger than a critical value $\varepsilon=2$, the Killing vector $\partial_{t}$ becomes spacelike for certain regions of $\theta$ with $\varepsilon>2$, which are so called as ergoregions. For $\varepsilon>2$, both solitons and black holes develop hair due to superradiance, and the spacetime with ergoregions in AdS may be unstable due to the superradiant scattering\cite{a14}. Furthermore, a family of deforming vacuum solutions with a noncompact, differential rotation boundary metric was numerically studied in\cite{a15}. In \cite{a16}, authors studied the influence of the hyperbolic and compact AdS black holes when the boundary metric changes by using AdS C-metric. One could construct the black holes with deforming horizon in $D=5$ minimal gauged supergravity by introducing the matter fields \cite{a17}. \\
\hspace*{0.6cm}We are interested in whether there are deforming solutions with the multipolar differential rotation boundary. In \cite{Li} authors try to numerically solve the Einstein equations and give a family of deforming solitons and black holes with even multipolar differential rotation boundary, which have the anti-symmetric rotation profile with respect to reflections on the equatorial plane and the total angular momentum of black hole is zero. We consider the configuration of quadrupolar rotation boundary in particular, and we obtain the numerical results of the deforming solitons and black holes. In quadrupolar differential rotation situation, solitons and black holes do not develop hair due to superradiance when the norm of Killing vector $\partial_{t}$ becomes spacelike for certain regions of $\theta$ with $\varepsilon\in(2,2.281)$. This is very different from the dipolar differential rotation situation. It could be observed that the black hole horizon is deformed into four hourglass shapes by isometric embedding the horizon. Furthermore, the numerical solutions of entropy and the quasinormal modes have analogous properties to that of dipolar differential rotation boundary \cite{r1}.\\
\hspace*{0.6cm}In this paper, we attempt to numerically solve the Einstein equation and give a family of deforming solitons and black holes with odd multipolar differential rotation boundary, which has the symmetric rotation profile with respect to reflections on the equatorial plane. Especially, considering the configuration of tripolar differential rotation boundary, we obtain the numerical results of the deforming solitons and black holes. Comparing with the results of dipolar differential rotation, we find that the norm of Killing vector $\partial_{t}$ becomes spacelike for certain regions of $\theta$ with $\varepsilon\in(2,2.124)$ at the temperature $T=1/\pi$. However, solitons and black holes with tripolar differential rotation do not develop hair due to superradiance, which is different from the case of dipolar rotation. Furthermore, we study the numerical solutions of entropy, and it is different from the cases of dipolar rotation and quadrupolar rotation when temperature increased, the maximal values of rotation parameter $\varepsilon$, below which the stable large black hole solutions could exist, are not a constant. When temperature is much higher than $T_{schw}=\sqrt{3}/2\pi\simeq0.2757$, which $T_{Schw}$ is the minimal temperature of AdS$_{4}$-Schwarzschild black hole, the norm of Killing vector $\partial_{t}$ keeps timelike for some regions of $\varepsilon<2$, solitons and black holes could be unstable and will develop hair due to superradiance. We find that the curves of entropy have different behaviours at different temperatures. When $T>T_{Schw}$, we first observe that the curves of entropy for large and small branches of black hole solutions separate into two curves. Then two branches of curves of entropy combine into one curve when $T=T_{Schw}$. When $T<T_{Schw}$, we again observe that the two curves of entropy separate into two curves, and each curve of entropy has two solutions at a fixed value of $\varepsilon$. In additions, by isometric embedding the horizon into hyperbolic space, the black hole horizon is deformed into three hourglass shapes. At last we study the quasinormal modes,  which have the analogous properties to that of dipolar and quadrupolar rotation boundary in \cite{Li,r1}.\\
\hspace*{0.6cm}The paper is organized as follows. In Sec. \ref{sec2}, we introduce the model of the deforming black holes with odd multipolar differential rotation boundary and the numerical DeTurck method. In Sec. \ref{sec3}, we construct a numerical models of soliton solutions with tripolar differential rotation boundary and show the numerical results of Kretschmann scalar and quasinormal modes. The numerical models and results of deforming black holes with tripoalr differential rotation boundary are shown in Sec. \ref{sec4}. The conclusions and discussions are given in the last section.
\section{Model and numerical method}\label{sec2}
\begin{figure}[H]
\centering
\includegraphics[width=0.32\textwidth]{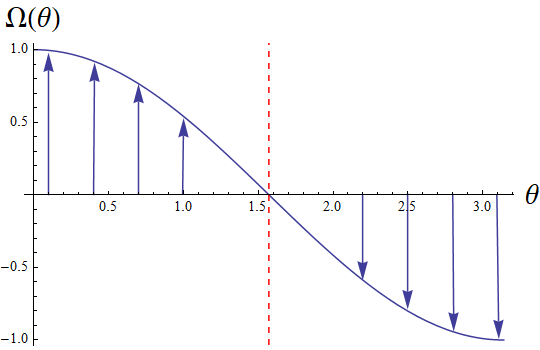}
\includegraphics[width=0.32\textwidth]{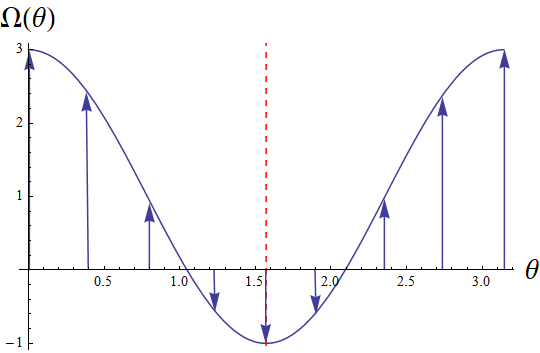}
\includegraphics[width=0.32\textwidth]{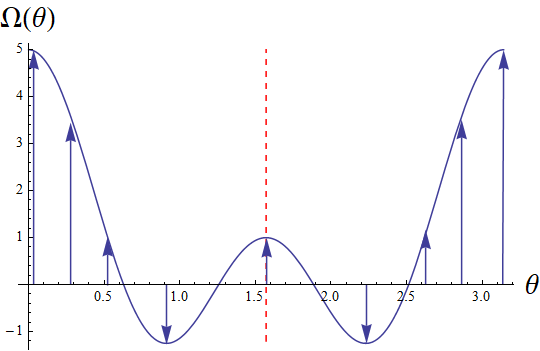}
\caption{The differential rotation profile $\Omega$ as a function of $\theta$ with dipolar rotation (left panel), tripolar rotation (middle panel) and pentapolar rotation (right panel), respectively. In all three graphs the arrow lines denote the orientation of differential rotation and the red dashed lines mean $\theta=\pi/2$.}
\label{fig1}
\end{figure}

\hspace*{0.6cm}We first begin with the model of the four-dimensional Einstein-Hilbert action with a negative cosmological constant $\Lambda$
\begin{equation}\label{eq2.1}
S=\frac{1}{16\pi G}\int d^{4}x\sqrt{-g}(R-2\Lambda),
\end{equation}
where $G$ is the gravitational constant, and the cosmological constant $\Lambda$ could be  represented by the AdS radius $L$ as $\Lambda=-3/L^{2}$, $g$ is the determinant of the metric tensor and $R$ is Ricci scalar. The equation of motion derived from (\ref{eq2.1}) takes the following form
\begin{equation}\label{eq2.2}
G_{ab}\equiv R_{ab}+\frac{3}{L^{2}}g_{ab}=0,
\end{equation}
The solution of Einstein equation (\ref{eq2.2}) is the well-known AdS-Schwarzschild black hole which described the static spherically symmetric black hole with mass. The metric is given by
\begin{equation}\label{eq2.3}
ds^{2}=-\left(1-\frac{2M}{r}+\frac{r^{2}}{L^{2}}\right)dt^{2}+\left(1-\frac{2M}{r}+\frac{r^{2}}{L^{2}}\right)^{-1}dr^{2}+r^{2}d\Omega^{2},
\end{equation}
where $d\Omega^{2}$ is the metric on the sphere $S^{2}$. The constant $M$ is the mass of black hole as measured from the infinite boundary. The horizon radius, denoted by $r_{+}$, and is the largest root of equation
\begin{equation}
1-\frac{2M}{r}+\frac{r^{2}}{L^{2}}=0.
\end{equation}
The Hawking temperature $T_{H}$ of AdS-Schwarzschild black hole is given by
\begin{equation}
T_{H}=\frac{L^{2}+3r_{+}^{2}}{4\pi L^{2}r_{+}}.
\end{equation}
As near the infinity, the metric (\ref{eq2.3}) is asymptotic to the anti-de Sitter spacetime, and the conformal boundary metric is given by
\begin{equation}
ds^{2}_{\partial}=r^{2}(-dt^{2}+d\theta^{2}+\sin^{2}\theta d\phi^{2}).
\end{equation}
In \cite{r1}, the authors added differential rotation to the boundary metric to obtain the new asymptotic anti-de Sitter solution, which is given by
\begin{equation}
ds^{2}_{\partial}=r^{2}(-dt^{2}+d\theta^{2}+\sin^{2}\theta [d\phi+\Omega(\theta)dt]^{2}),
\end{equation}
where $\Omega(\theta)=\varepsilon\cos\theta$ is the dipolar differential rotation. The norm of Killing vector $\partial_{t}$ is
\begin{equation}
||\partial_{t}||^{2}=-1+\frac{\varepsilon^{2}}{4}\sin^{2}(2\theta),
\end{equation}
and the maximum value is at $\theta=\pi/4$.\\
\hspace*{0.6cm} In order to construct higher odd multipolar differential rotation of the conformal boundary, we  could take the following form of  Killing vector $ \partial_{t}$
\begin{equation}\label{eq2.9}
\begin{aligned}
\| \partial_{t}\|^{2}=-1+\frac{\varepsilon^{2}}{4}\sin^{2}(k\theta),\quad k=3,5,7,\cdot\cdot\cdot,
\end{aligned}
\end{equation}
which corresponds to the odd multipole differential rotation
\begin{equation}\label{eq2.10}
\Omega(\theta)=\left\{
\begin{aligned}
2\varepsilon[1+2\cos(2\theta)],\quad &k=3,\\
2\varepsilon[\csc(\theta)\sin(5\theta)],\quad &k=5,\\
2\varepsilon[\csc(\theta)\sin(7\theta)],\quad &k=7,\\
 \cdots,\quad\quad\quad\quad &k=\cdots.
\end{aligned}
\right.
\end{equation}
 For the case of
$k=3$, the $\|\partial_{t}\|^{2}$ has maximal value at $\theta=\pi/6$,  and when $\varepsilon>2$ in the certain regions of $\theta$, the Killing vector $\partial_{t}$ becomes spacelike. In Fig.$\ $\ref{fig1}, we draw the graphs of differential rotation $\Omega$ as the functions of $\theta$ with dipolar rotation (left panel), tripolar rotation (middle panel) and pentapolar rotation (right panel), respectively. In all three graphs the arrow lines denote the orientation of differential rotation. The dipolar differential rotation $\Omega$ is the anti-symmetric function with respect to reflections on the equatorial plane $\theta=\pi/2$, and the total angular momentum of black hole is zero. But in the odd polar situation, $\Omega$ is symmetric function with respect to reflections on the equatorial plane $\theta=\pi/2$.
We would like to use DeTurck method\cite{r4,r5,r6} to solve Einstein equation (\ref{eq2.2}) to obtain the numerical solutions. The DeTurck method is to add a gauge fixing term to Einstein equation, which is known as Einstein-DeTurck equation
\begin{equation}\label{eq2.11}
R_{ab}+\frac{3}{L^{2}}g_{ab}-\nabla(_{a}\xi_{b})=0,
\end{equation}
where $\xi^{a}=g^{bc}(\Gamma^{a}_{bc}[g]-\Gamma^{a}_{bc}[\widetilde{g}])$ is the Levi-Civita connection and $\widetilde{g}$ is an appropriate reference metric of our choice. It is noted that the reference metric $\widetilde{g}$ should be chosen to have the same boundary and horizon structure as $g$. When $\xi=0$, the equation (\ref{eq2.11}) recovers to the Einstein equation.\\
\hspace*{0.6cm}By using numerical methods to solve these equations of motion, we obtain two classes of solutions, which are horizonless soliton solutions with $r_{H}=0$ and black hole solutions with $r_{H}>0$. The soliton solutions could be seen as deformations of the global anti-de Sitter spacetime, and the black hole solutions  correspond to deformations of AdS$_{4}$ black holes. For simplicity, in our paper we only show the numerical results of the tripolar differential rotation ($k=3$), and the situations of higher odd-multipole differential rotation have similar behaviours as that of solutions of tripolar differential rotation.\\
\hspace*{0.6cm}We use the finite element methods in the integration regions $0\leqslant x\leqslant1$ and $0 \leqslant y \leqslant 1$ defined on non-uniform grids, allowing the grids to be more finer grid points near the boundaries of $y=0$ and $y=1$. Our iterative process is the Newton-Raphson method. The relative error for the numerical solutions in this work is estimated to be below $10^{-6}$. In order to keep good agreement with the aforementioned error, the grid size has to be increased and typically a $120\times200$ to $120 \times 370$ grid was used. In all the calculations and plots, we set $L=1$.
\section{Soliton solutions}\label{sec3}
\begin{figure}[t]
\centering
\begin{minipage}[c]{0.5\textwidth}
\centering
\includegraphics[scale=0.25]{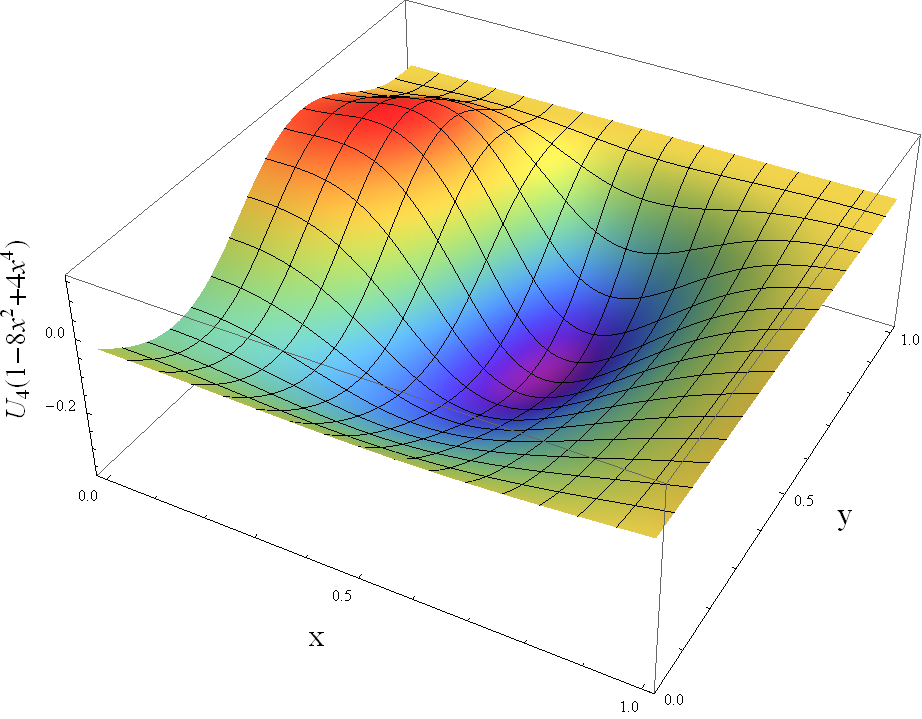}
\end{minipage}%
\begin{minipage}[c]{0.5\textwidth}
\centering
\includegraphics[scale=0.25]{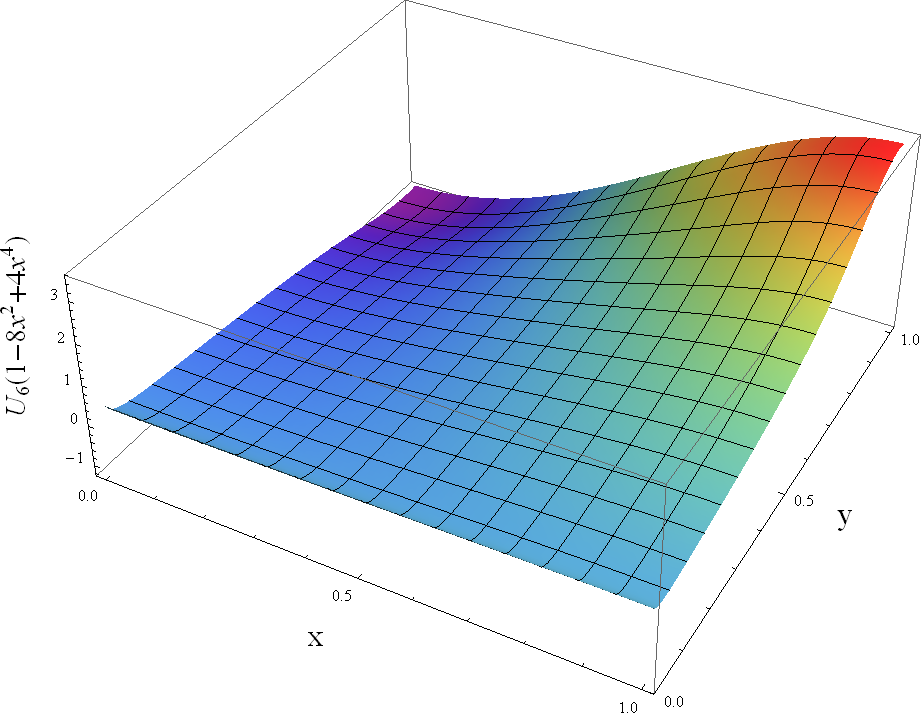}
\end{minipage}
\caption{The left panel shows $U_{4}(1-8x^{2}+4x^{4})$ and the right panel shows $U_{6}(1-8x^{2}+4x^{4})$ of soliton solution with the tripolar boundary rotation. Two solutions have $\varepsilon=2.3$.}
\label{fig2}
\end{figure}

\hspace*{0.6cm}We make a coordinate transformation that the variable $y$ is related to the AdS$_{4}$ radial coordinate via $r=Ly\sqrt{2-y^{2}}/(1-y^{2})$, and $x$ is the variable about the standard polar angle on $S^{2}$ by $\sin\theta=1-x^{2}$, which implies that the new radial coordinate $y\in[0, 1]$ and polar angle coordinate $x\in[0, 1]$. Thus the inner and outer boundaries of the shell are fixed at $y = 0$ and $y = 1$, respectively. In order to solve Einstein equation (\ref{eq2.2}) numerically with a tripolar differential rotation (\ref{eq2.10}), we introduce the following ansatz of solitonic solutions
\begin{equation}
\begin{aligned}
ds^{2}=\frac{L^{2}}{(1-y^{2})^{2}}\bigg\{-U_{1}\,dt^{2}+\frac{4\,U_{2}}{2-y^{2}}\,dy^{2}&+y^{2}(2-y^{2})\bigg[\frac{4\,U_{3}}{2-x^{2}}\left(dx+\frac{1-8x^{2}+4x^{4}}{y}\,U_{4}\,dy\right)^{2}\\
&+(1-x^{2})^{2}\,U_{5}\,\left(d\phi+y\,(1-8x^{2}+4x^{4})\,U_{6}\,dt\right)^{2}\bigg]\bigg\}.
\end{aligned}
\end{equation}
In $x\in[0, 1]$, the zero value of $\Omega(\theta)$ are appeared at $x=\frac{1}{2}(-1+\sqrt{3})$, and this will cause difficulties in numerical calculation when we solve the Einstein-DeTurck equations. To solve these difficulties, we combine the term $1-8x^{2}+4x^{4}$ and the functions $U_{i}$ $(i=4,6)$ into new functions $\tilde{U}_{i}$ $(i=4,6)$. The ansatz with $\tilde{U}_{4}=U_{4}(1-8x^{2}+4x^{4})$ and $\tilde{U}_{6}=U_{6}(1-8x^{2}+4x^{4})$ could be written as
\begin{equation}\label{eq3.2}
\begin{aligned}
ds^{2}=\frac{L^{2}}{(1-y^{2})^{2}}\bigg\{-U_{1}\,dt^{2}+\frac{4\,U_{2}}{2-y^{2}}\,dy^{2}&+y^{2}(2-y^{2})\bigg[\frac{4\,U_{3}}{2-x^{2}}\left(dx+\frac{1}{y}\,\tilde{U}_{4}\,dy\right)^{2}\\
&+(1-x^{2})^{2}\,U_{5}\,\left(d\phi+y\,\tilde{U}_{6}\,dt\right)^{2}\bigg]\bigg\},
\end{aligned}
\end{equation}
where $U_{i}$, $(i=1,2,3,5)$ and $\tilde{U}_{4}$, $\tilde{U}_{6}$ are the functions of $(x,y)$.\\
\hspace*{0.6cm}We need to obtain the asymptotic behaviours of functions $U_{i}$ $(i=1,2,3,5)$ and $\tilde{U}_{4}$, $\tilde{U}_{6}$ before numerically solve the partial differential equations, which are also the boundary conditions we need. Because the solutions have properties of polar angle reflection symmetry $\theta\rightarrow\pi-\theta$ on the equatorial plane, so we could consider the coordinate range $\theta\in[0,\pi/2]$, i.e $x\in[0,1]$. Therefore, we require the functions to satisfy the following Neumann boundary conditions on the equatorial plane $x=0$
\begin{equation}
\left\{
\begin{aligned}
&\partial_{x}U_{i}(0,y)=0,\quad i=1,2,3,5\\
&\partial_{x}\tilde{U}_{i}(0,y)=0, \quad i=4,6.
\end{aligned}
\right.
\end{equation}
In addition, we impose the Dirichlet boundary conditions on $\tilde{U}_{4}$ at $x=1$
\begin{equation}
\tilde{U}_{4}(1,y)=0,
\end{equation}
and the Neumann boundary conditions on the other functions
\begin{equation}
\partial_{x}U_{1}(1,y)=\partial_{x}U_{2}=\partial_{x}U_{3}=\partial_{x}U_{5}=\partial_{x}\tilde{U}_{6}=0.
\end{equation}
Moreover, we obtain the condition $U_{3}(1,y)=U_{5}(1,y)$ by expanding the equations of motion near $x=1$, and the asymptotic behaviours near the conformal boundary $y=1$ are
\begin{equation}
\begin{aligned}
U_{1}(x,1)=U_{2}(x,1)=U_{3}(x,1)=U_{5}(x,1),\\
\tilde{U}_{4}=0,\quad \tilde{U}_{6}=(1-8x^{2}+4x^{4})\varepsilon.
\end{aligned}
\end{equation}
Finally, by expanding the equations of motion near $y=0$ as a power series in $y$, we have
\begin{equation}
\left\{
\begin{aligned}
&\partial_{y}U_{i}(x,0)=0,\quad i=1,2,3,5,\\
&\partial_{y}\tilde{U}_{i}(x,0)=0, \quad i=4,6.
\end{aligned}
\right.
\end{equation}
In addition, we could choose the reference metric $\tilde{g}$ which is given by metric (3.2) with $U_{1}=U_{2}=U_{3}=U_{5}=1$, $\tilde{U}_{4}=0$ and $\tilde{U}_{6}=\varepsilon(1-8x^{2}+4x^{4})$.\\
\hspace*{0.6cm}In Fig.$\ $\ref{fig2}, we show the typical soliton results of our numerical code for $\tilde{U}_{4}$ in the left panel and $\tilde{U}_{6}$ in the right panel with tripolar boundary rotation, and all the two figures have same parameter $\varepsilon=2.3$. According to the numerical results, we find there exists stationary axisymmetric soliton solutions for $\varepsilon<\varepsilon_{c}=2.5264$, where $\varepsilon_{c}$ is the maximal value. Furthermore, we find the soliton solutions for each value of $\varepsilon\in(2.5204,2.5264)$ have two branches.
\subsection{Kretschman scalar}
\hspace*{0.6cm}It is very important to know whether the spacetime of solution is regular or not when we obtain a solution of Einstein equation. In general, the Ricci scalar is the simplest curvature invariant of a Riemannian manifold. However, in our model the Ricci tensor is $R=2\Lambda$. The another choice is to check the behaviour of the Kretschmann scalar which is invariant and it could indicate the flatness of a chosen manifold. The Kretschmann scalar is written as
\begin{equation}
K=R_{\alpha\beta\gamma\delta}R^{\alpha\beta\gamma\delta},
\end{equation}
where $R_{\alpha\beta\gamma\delta}$ is the Riemann curvature tensor and because it is a sum of squares of tensor components, Kretschmann scalar is a quadratic invariant.\\
\begin{figure}[t]
\centering
\includegraphics[scale=0.3]{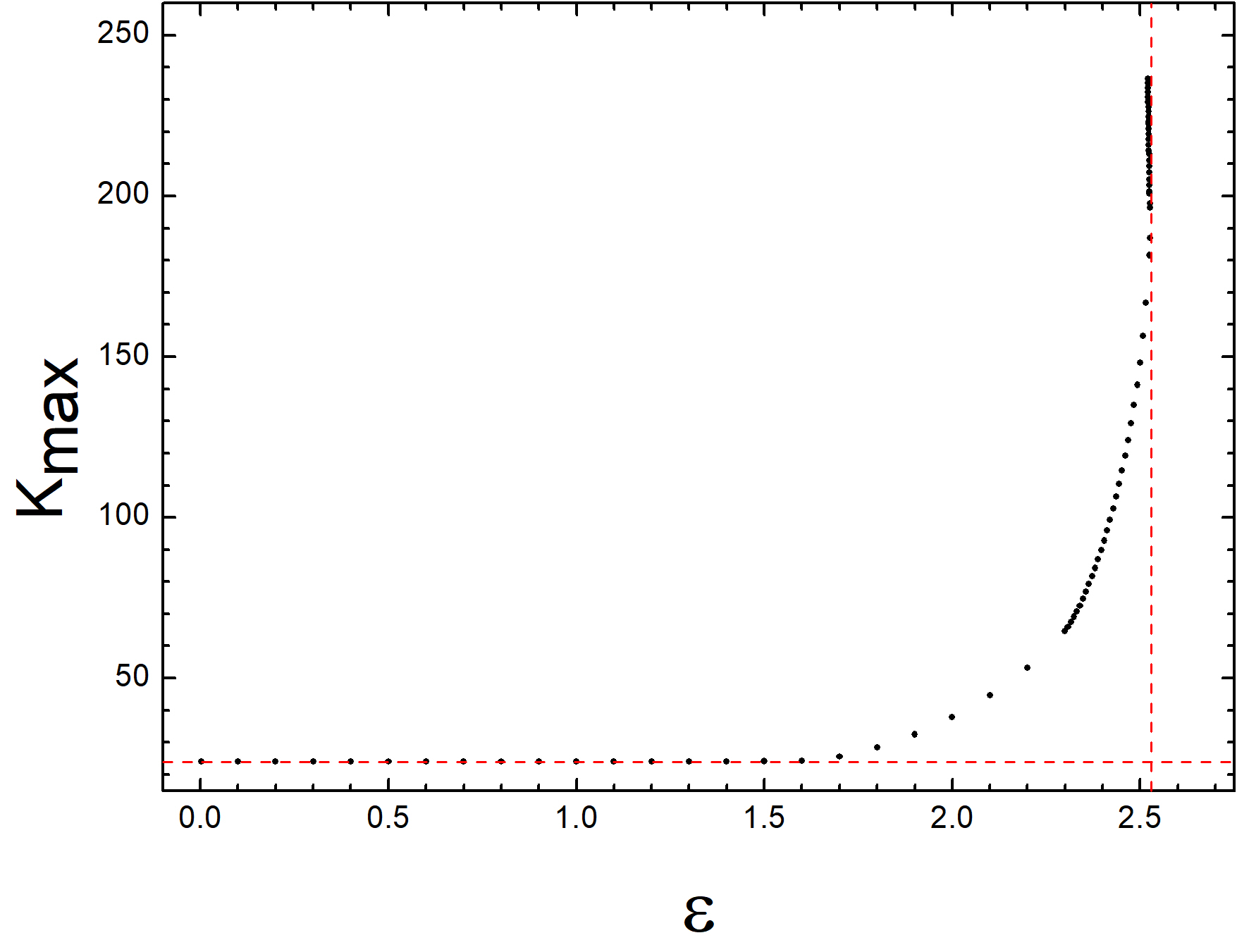}
\caption{The maximum of Kretschmann scalar as a function of the boundary rotation parameter $\varepsilon$ for the soliton solutions. The vertical red dashed line marks the max $K=24/L^{2}$ for AdS$_{4}$ spacetime, and the horizontal red dashed line marks the $\varepsilon_{c}=2.5264$.}
\label{fig3}
\end{figure}
\hspace*{0.6cm}We present the maximal value of Kretschmann scalar as the function of $\varepsilon$ in Fig.$\ $\ref{fig3}, it could be seen that the maximal value of Kretschmann scalar increases with the increase of $\varepsilon$. When the rotation parameter $\varepsilon$ approaches zero, the maximal value of $K$ is equal to maximal value of $K$ for pure AdS$_{4}$. Furthermore, for $\varepsilon\in(2.5204,\varepsilon_{c})$, we find that there are two soliton solutions exist for a fixed value of $\varepsilon$, but it is too difficult for our numerical method to calculate the maximal value for the second branch of Kretschmann scalar $K$ when $\varepsilon<2.5204$.
\subsection{The quasi-normal modes}
\hspace*{0.6cm}In order to study the linear stability of soliton solutions with tripolar boundary rotation, we will study the quasinormal modes (QNMs), which are characteristics of the background spacetime. Following the method in papers \cite{r1,d1,d2}, we consider a free, massless scalar field, obeying a massless Klein-Gordon equation
\begin{equation}
\nabla^{2}\Psi=0,
\end{equation}
and the scalar field could be separated into the standard form
\begin{equation}
\Psi=\hat{\Psi}_{\omega,m}(x,y)e^{-i\omega t+im\psi},\quad m=\pm1,\pm2,...,
\end{equation}
where $\omega$ is the frequency of the complex scalar field and $m$ is the azimuthal harmonic index. On the soliton background (\ref{eq3.2}), the scalar field could be separated into
\begin{equation}
\Psi(t,x,y,\psi)=e^{i(m\psi-\omega t)}y^{|m|}(1-y^{2})^{3}(1-x^{2})^{|m|}\xi(x,y).
\end{equation}
The boundary conditions are imposed as follow: at $x=\pm1$, we require $\partial_{x}\xi(x,y)=0$; at $y=0$, we require $\partial_{y}\xi(x,y)=0$, and at $y=1$, we require
\begin{equation}
\partial_{y}\xi(x,y)=-|m|\xi(x,y).
\end{equation}
\begin{figure}[t]
\centering
\includegraphics[scale=0.3]{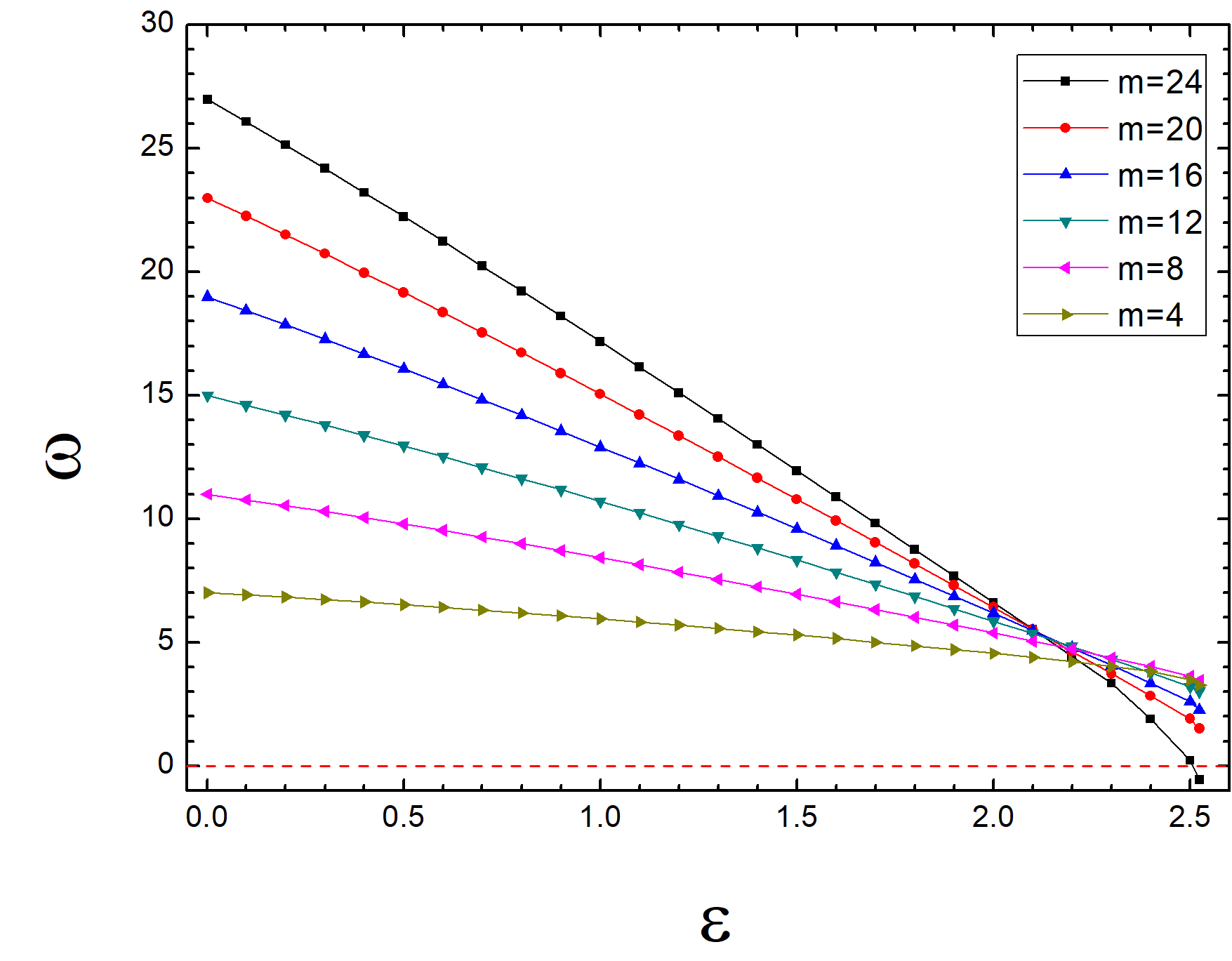}
\caption{Normal mode frequencies against the rotation parameter $\varepsilon$ for the soliton metric. The red dashed line shows the $\omega=0 $.}
\label{fig4}
\end{figure}
\hspace*{0.6cm}In Fig.$\ $\ref{fig4}, we plot the normal mode frequencies $\omega$ as the functions of the rotation parameter $\varepsilon$ for corresponding values of $m$. We could see that the frequency $\omega$ becomes negative at a specific value of $\varepsilon$ when $m=24$, and the frequencies are always positive modes in the spectrum of perturbations when $m\leqslant23$. When $m\geqslant24$, one could expect some branches of soliton with scalar hair $\Psi$ condensation could be found.
\section{Black Hole solutions}\label{sec4}
\hspace*{0.6cm}In order to obtain the black hole solutions with the tripolar differential rotation, we consider the ansatz of metric
\begin{equation}
\begin{aligned}
ds^{2}=\frac{L^{2}}{(1-y^{2})^{2}}\bigg\{-y^{2}\,\tilde{\Gamma}(y)\,U_{1}\,dt^{2}+&\frac{4y_{p}^{2}\,U_{2}\,dy^{2}}{\tilde{\Gamma}(y)}
+y_{p}^{2}\bigg[\frac{4U_{3}}{2-x^{2}}\left(dx+(1-8x^{2}+4x^{4})\,y\,U_{4}\,dy\right)^{2}\\
&+(1-x^{2})^{2}\,U_{5}\left(d\phi+y^{2}\,(1-8x^{2}+4x^{4})\,U_{6}\,dt\right)^{2}\bigg]\bigg\}
\end{aligned}
\end{equation}
Considering the same difficulties in numerical calculation, which are similar to the case of the soliton, we obtain the ansatz with $\tilde{U}_{4}=U_{4}(1-8x^{2}+4x^{4})$ and $\tilde{U}_{6}=U_{6}(1-8x^{2}+4x^{4})$ as follow
\begin{subequations}
\begin{equation}\label{eq4.2}
\begin{aligned}
ds^{2}=\frac{L^{2}}{(1-y^{2})^{2}}\bigg\{-y^{2}\,\tilde{\Gamma}(y)\,U_{1}\,dt^{2}&+\frac{4y_{p}^{2}\,U_{2}\,dy^{2}}{\tilde{\Gamma}(y)}
+y_{p}^{2}\bigg[\frac{4U_{3}}{2-x^{2}}\left(dx+y\,\tilde{U}_{4}\,dy\right)^{2}\\
&+(1-x^{2})^{2}\,U_{5}\left(d\phi+y^{2}\,\tilde{U}_{6}\,dt\right)^{2}\bigg]\bigg\},
\end{aligned}
\end{equation}
where
\begin{equation}
\begin{aligned}
\tilde{\Gamma}(y)&=\Gamma(y)\sigma+y_{p}^{2}(1-\sigma),\quad \Gamma(y)&=(1-y^{2})^{2}+y_{p}^{2}(3-3y^{2}+y^{4}),
\end{aligned}
\end{equation}
\end{subequations}
with $U_{i}$ $(i=1,2,3,5)$ and $\tilde{U}_{4}$, $\tilde{U}_{6}$ are the functions of $(x,y)$. The line element (\ref{eq4.2}) will reduce to AdS$_4$ black hole in global coordinates when $U_{1}=U_{2}=U_{3}=U_{5}=\sigma=0$, and $U_{4}=U_{6}=0$. The variable $y$ is related to the usual radial coordinate via $r=L\,y_{p}/(1-y^{2})$, and $x$ is the variable about standard polar angle on $S^{2}$ via $\sin\theta=1-x^{2}$.\\
\begin{figure}[t]
\centering
\begin{minipage}[c]{0.5\textwidth}
\centering
\includegraphics[scale=0.25]{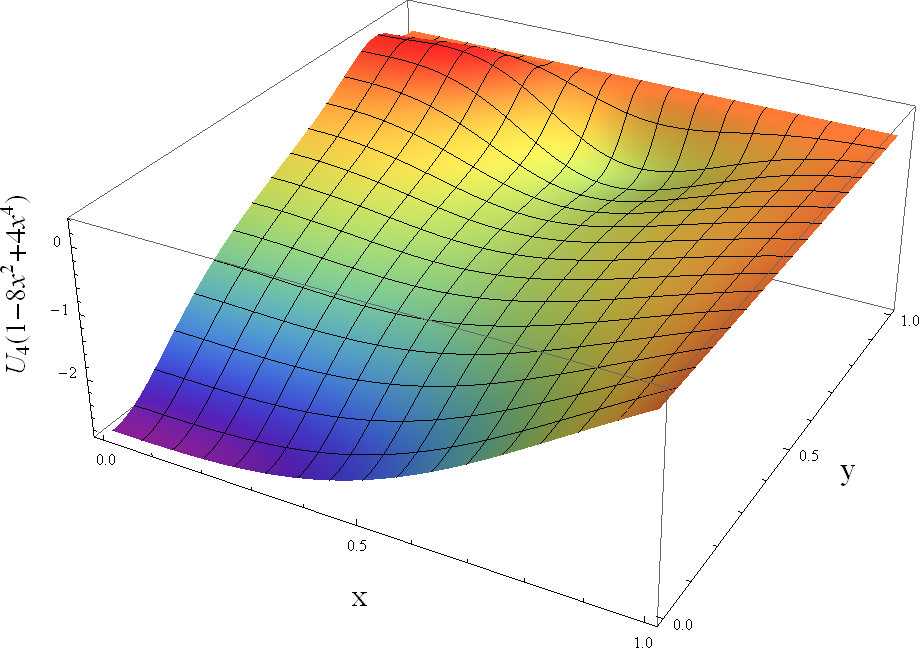}
\end{minipage}%
\begin{minipage}[c]{0.5\textwidth}
\centering
\includegraphics[scale=0.25]{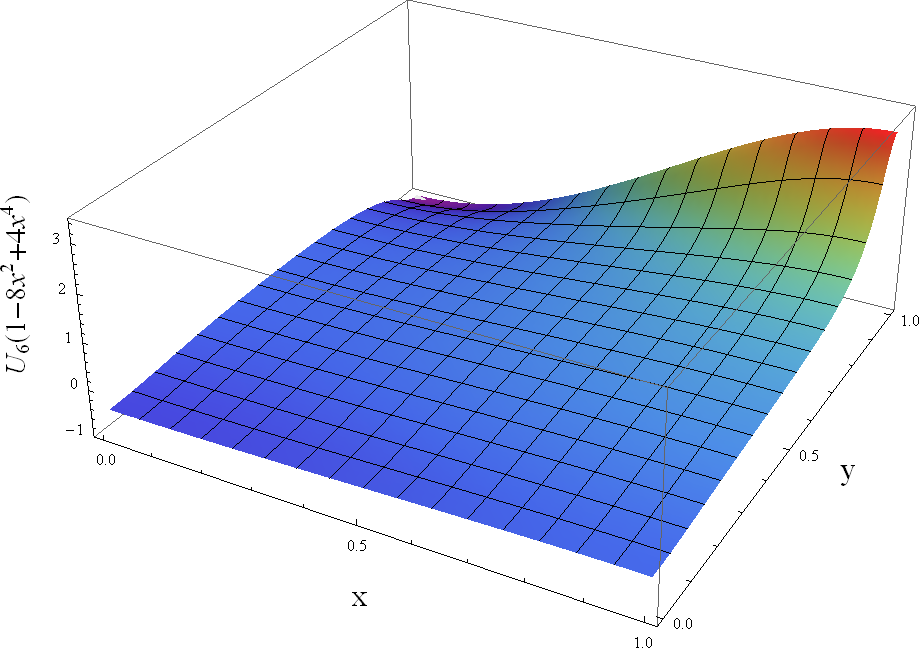}
\end{minipage}
\begin{minipage}[c]{0.5\textwidth}
\centering
\includegraphics[scale=0.25]{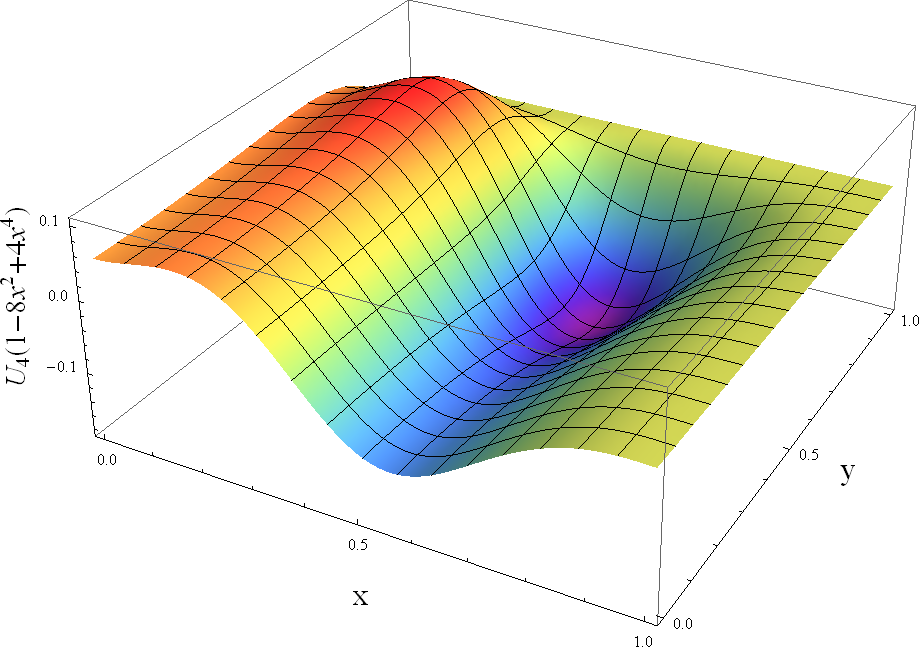}
\end{minipage}%
\begin{minipage}[c]{0.5\textwidth}
\centering
\includegraphics[scale=0.25]{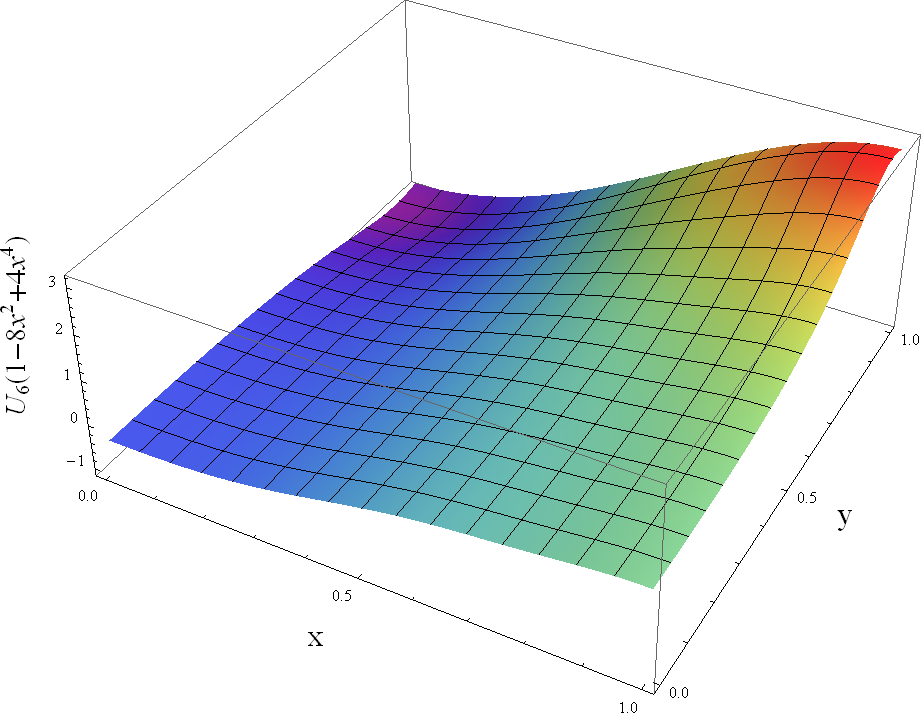}
\end{minipage}
\caption{\emph{Top}: The distributions of $\tilde{U}_{4}$ and $\tilde{U}_{6}$ as functions of $x$ and $y$ for small black hole. \emph{Bottom}: The distributions of $\tilde{U}_{4}$ and $\tilde{U}_{6}$ as functions of $x$ and $y$ for large black hole. All the solutions have the rotation parameter $\varepsilon=2$.}
\label{fig5}
\end{figure}
\hspace*{0.6cm}In the ansatz (\ref{eq4.2}) we have three parameters: $\varepsilon$ is the amplitude of boundary rotation, and $(y_{p},\sigma)$ are  related to  the black hole temperature. The Hawking temperature is computed in the usual way by metric:
\begin{equation}
T=\frac{1}{4\pi}\sqrt{-g^{tt}g^{\alpha\beta}\partial_{\alpha}g_{tt}\partial_{\beta}g_{tt}}|_{r=r_{H}}=\frac{2\sigma y_{p}^{2}+y_{p}^{2}+\sigma}{4\pi y_{p}}.
\end{equation}
The temperature has a minimal value at $y_{p}=1/\sqrt{3}$ when $\sigma=1$, coinciding with the minimal temperature of a Schwarzschild-AdS$_{4}$, occurring at $T_{Schw}=\sqrt{3}/(2\pi)\approx0.2757$. Usually, we will have two branches of solutions at the same temperature with a fixed value of $\delta$, one is called \emph{large} black holes with larger $y_{p}$ and another is \emph{small} black holes with smaller $y_{p}$.\\
\hspace*{0.6cm}The boundary conditions are similar to the soliton case. At $x=0$ and $y=0$, the functions $U_{i}$ satisfy the Neumann conditions
\begin{equation}
\left\{
\begin{aligned}
&\partial_{x}U_{i}(0,y)=\partial_{y}U_{i}(x,0)=0, \quad i=1,2,3,5,\\
&\partial_{x}\tilde{U}_i(0,y)=\partial_{y}\tilde{U}_i(x,0)=0, \quad i=4,6,
\end{aligned}
\right.
\end{equation}
at $x=1$, we choose $\tilde{U}_{4}=0$, $U_{3}=U_{5}$, and $\partial_{x}U_{1}=\partial_{x}U_{2}=\partial_{x}U_{3}=\partial_{x}U_{5}=\partial_{x}\tilde{U}_{6}=0$. At $y=1$, we set $\tilde{U}_4=0$, $\tilde{U}_{6}=\varepsilon(1-8x^{2}+4x^{4})$ and $U_{1}=U_{2}=U_{3}=U_{5}=1$. Furthermore, we could obtain $U_{1}(x,0)=U_{2}(x,0)$ by expanding the equations of motion near $y=0$. The reference metric $\tilde{g}$ is given by line element (\ref{eq4.2}) with $U_{1}=U_{2}=U_{3}=U_{5}=1$, $\tilde{U}_{4}=0$ and $\tilde{U}_{6}=\varepsilon(1-8x^{2}+4x^{4})$.\\
\hspace*{0.6cm}We present the typical results of $\tilde{U}_{4}$ and $\tilde{U}_{6}$ with tripolar differential rotation for small and large black hole solutions in Fig.$\ $\ref{fig5}. The left column exhibits the distributions of $\tilde{U}_{4}$ for small black hole (top) and large black hole (bottom), and the right column exhibits the distributions of $\tilde{U}_{6}$ for small black hole (top) and large black hole (bottom). We could observe that the deformations of $\tilde{U}_4$ and $\tilde{U}_6$ for the case of large black hole are larger than the case of small black hole.

\subsection{Entropy}
\begin{figure}[t]
\centering
\begin{minipage}[c]{0.5\textwidth}
\centering
\includegraphics[scale=0.25]{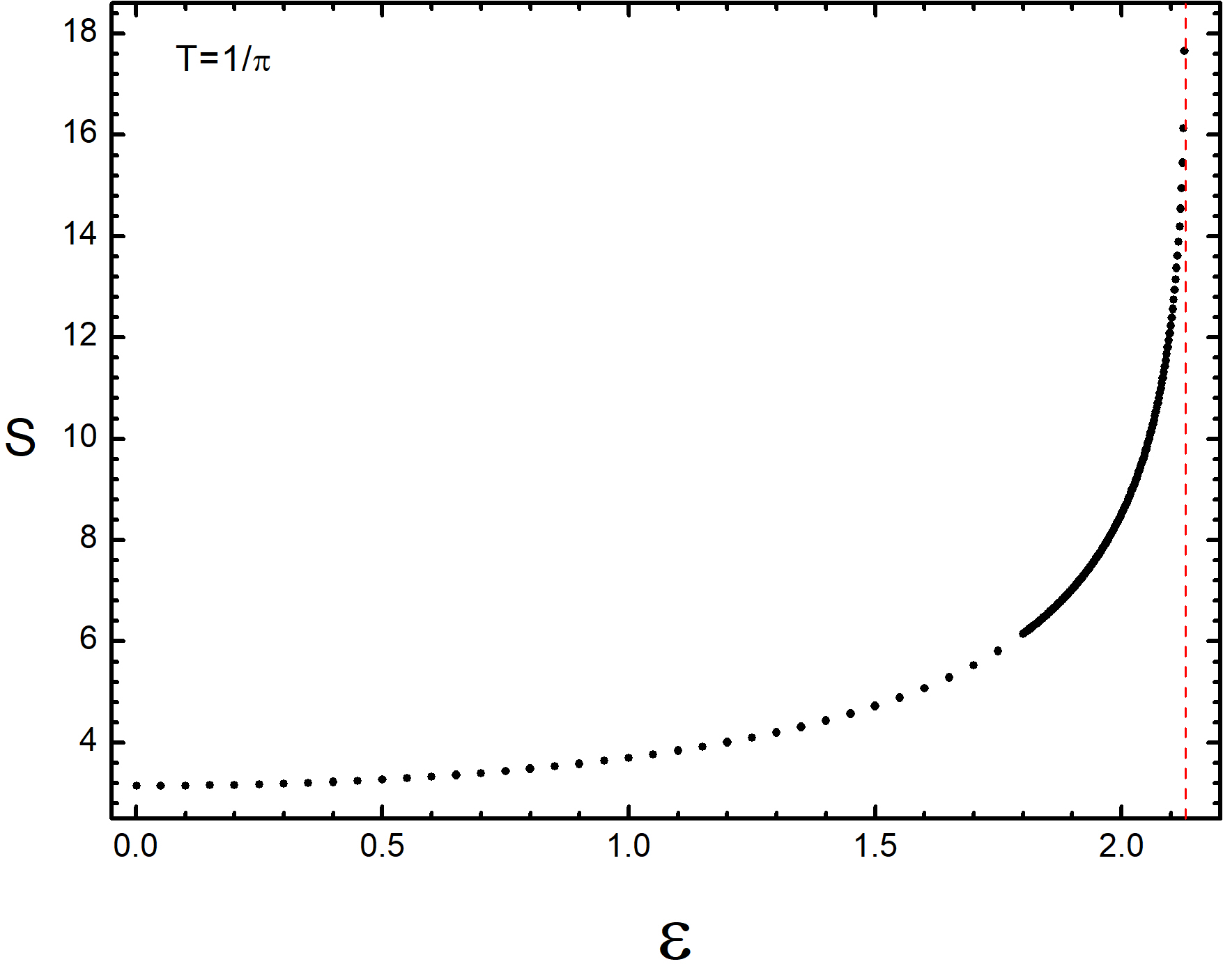}
\end{minipage}%
\begin{minipage}[c]{0.5\textwidth}
\centering
\includegraphics[scale=0.257]{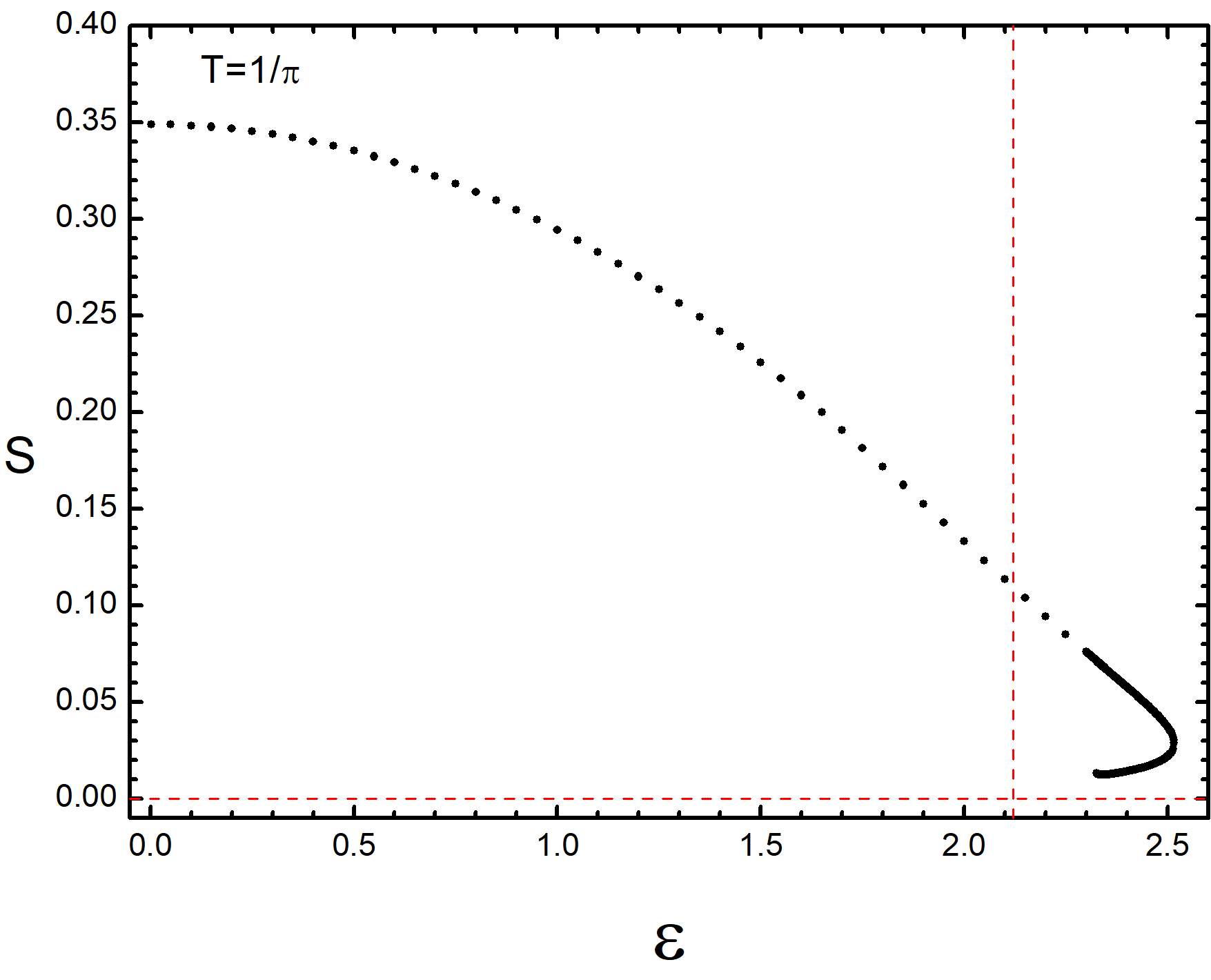}
\end{minipage}
\caption{\emph{Left}: The entropy against boundary rotation parameter $\varepsilon$ for the large branch of black hole solutions with a temperature $T = 1/\pi$. \emph{Right}: The entropy against boundary rotation parameter $\varepsilon$ for small branch of black hole solutions with same temperature. The vertical red dashed lines in the left and right show $\varepsilon=2.124$ and horizontal red dashed line shows $S=0$.}
\label{fige10}
\end{figure}

\hspace*{0.6cm}In this subsection we discuss the entropy of deforming black holes with tripolar differential rotation. The entropy associated with the black hole horizon is given by
\begin{equation}
S=\frac{A}{4G_{N}}=\frac{2\pi\,y_{p}^{2}\,L^{2}}{G_{N}}\int_{0}^{1}dx\frac{1-x^{2}}{\sqrt{2-x^{2}}}\sqrt{U_{3}(x,0)U_{5}(x,0)}.
\end{equation}
\hspace*{0.6cm}We present our numerical results of entropy for large and small branches of black hole solutions with temperature $T=1/\pi$ in Fig.$\ $\ref{fige10}, where the vertical red dashed line represents $\varepsilon=2.124$.
The entropy for the large branch of black hole solutions with $y_{p}=1$ is shown in the left panel of Fig.$\ $\ref{fige10}, we could see that the entropy always increases with the increase of $\varepsilon$, and the axially symmetric black hole solutions with tripolar differential rotation couldn't be found when $\varepsilon>2.124$. The right panel of Fig.$\ $\ref{fige10} shows the entropy for the small branch of black hole solutions with $y_{p}=1/3$. The curve of entropy for small branch of black hole solutions decreases with the increase of rotation parameter $\varepsilon$, and then it reaches the minimal value at $\varepsilon_{c}=2.5264$. Further decreasing $\varepsilon$, we obtain the second set of solution with lower entropy and the entropy decreases with the decrease of $\varepsilon$.\\
\begin{figure}[t]
\centering
\includegraphics[scale=0.3]{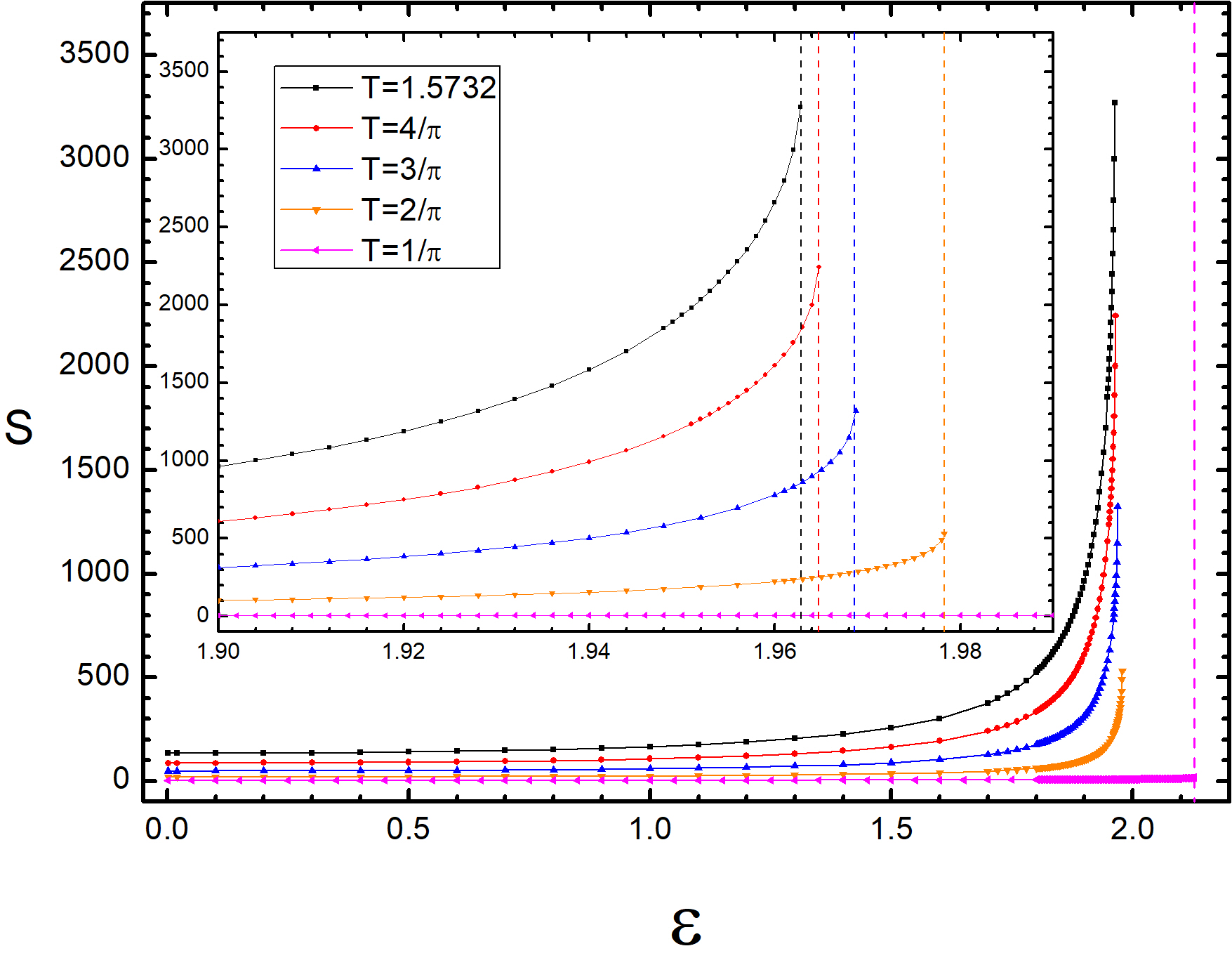}
\caption{The entropy against boundary rotation parameter $\varepsilon$ for large black holes at $T\in(1/\pi,1.5732)$. The inset shows the asymptotic behaviours of curves of entropy at different temperatures $T\in(2/\pi,1.5723)$ around the limits. The vertical dashed lines of different colors represent the limits of different temperatures. The purple line represents the entropy of large black hole at $T=1/\pi$. Note that because of the scale of axis, it looks like a straight line. The details of entropy at $T=1/\pi$ are shown in Fig.$\ $\ref{fige10}.}
\label{fig9}
\end{figure}
\hspace*{0.6cm}We find that in the case of tripolar differential rotation, the norm of Killing vector $\partial_{t}$ in Eq. (\ref{eq2.9}) becomes spacelike for the certain regions of $\theta$ with $\varepsilon\in(2,2.124)$. However, solitons and black holes with tripolar differential rotation still exist and do not develop hair due to superradiance, which is different from the case of dipolar differential rotation. Furthermore, we could obtain the entropy of large black holes at high temperature $(T\gg T_{schw})$, and we find that
the maximum values of $\varepsilon$, below which the stable large black hole solutions could exist, are not a constant under the different high temperatures.
With the increase of temperature $T$, the maximum value of $\varepsilon$ decreases, which is different from the case in \cite{Li,r1}. We show our results of entropy for large black holes at different high temperatures in Fig.$\ $\ref{fig9}.\\
\hspace*{0.6cm}In Fig.$\ $\ref{fig9}, the lines of different colors represent the curves of entropy of large black holes for different high temperatures with $T\in(1/\pi,1.5732)$, and the purple vertical dashed line indicates the maximum value of $\varepsilon$ at $T=1/\pi$. The inset  of Fig.$\ $\ref{fig9} shows the asymptotic behaviours at the maximum values of $\varepsilon$ for $T\in(2/\pi,1.5732)$, and the vertical dashed lines of different colors represent the maximum values of $\varepsilon$ at different temperatures. From the inset, it is clearly that with the increase of temperature $T$, the maximum value of $\varepsilon$ decreases.  It is interesting that
when temperature is much higher than $T_{schw}$,  the maximum value of $\varepsilon$  could lower than $\varepsilon=2$, which is the critical value for the Killing vector $\partial_{t}$ becomes spacelike for certain regions of $\theta$.  This means that even though  the norm of Killing vector $\partial_{t}$
keeps timelike for the some regions of $\varepsilon<2$, the solitons and black holes with tripolar differential rotation  could still develop hair due to superradiance at high temperature.\\
\hspace*{0.6cm}For the low temperature $T\leqslant T_{Schw}=\sqrt{3}/2\pi\simeq0.2757$ situation, we present our results in Fig.$\ $\ref{fige13}. The dashed lines with hollow symbols represent the entropy for large branch of black hole solutions and the solid lines with solid symbols represent the entropy for small branch of black hole solutions. Two types of lines of same color denote entropy for large and small branches of black hole solutions respectively at same temperature.
\begin{figure}[t]
\centering
\begin{minipage}[c]{0.5\textwidth}
\centering
\includegraphics[scale=0.255]{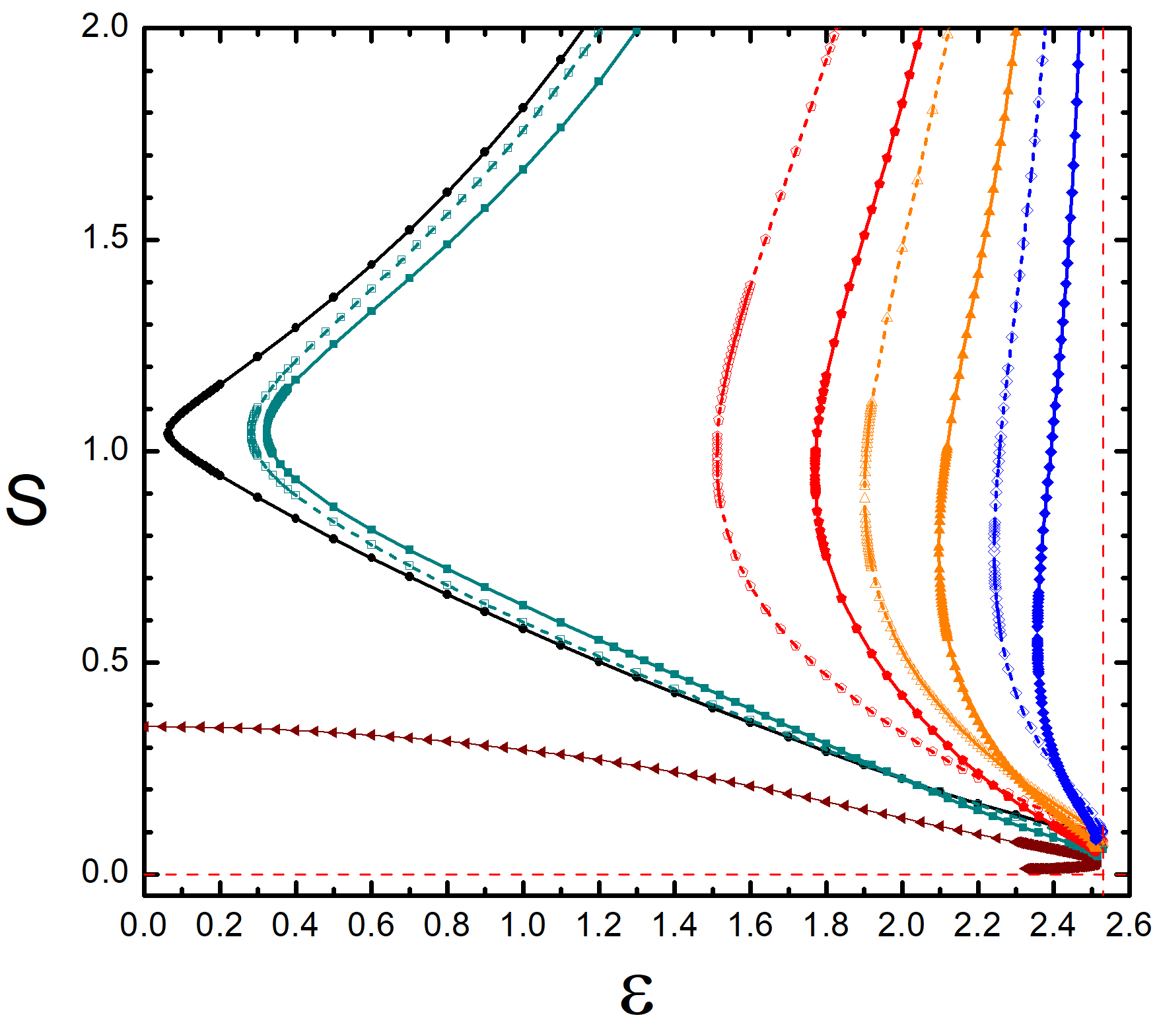}
\end{minipage}%
\begin{minipage}[c]{0.5\textwidth}
\centering
\includegraphics[scale=0.245]{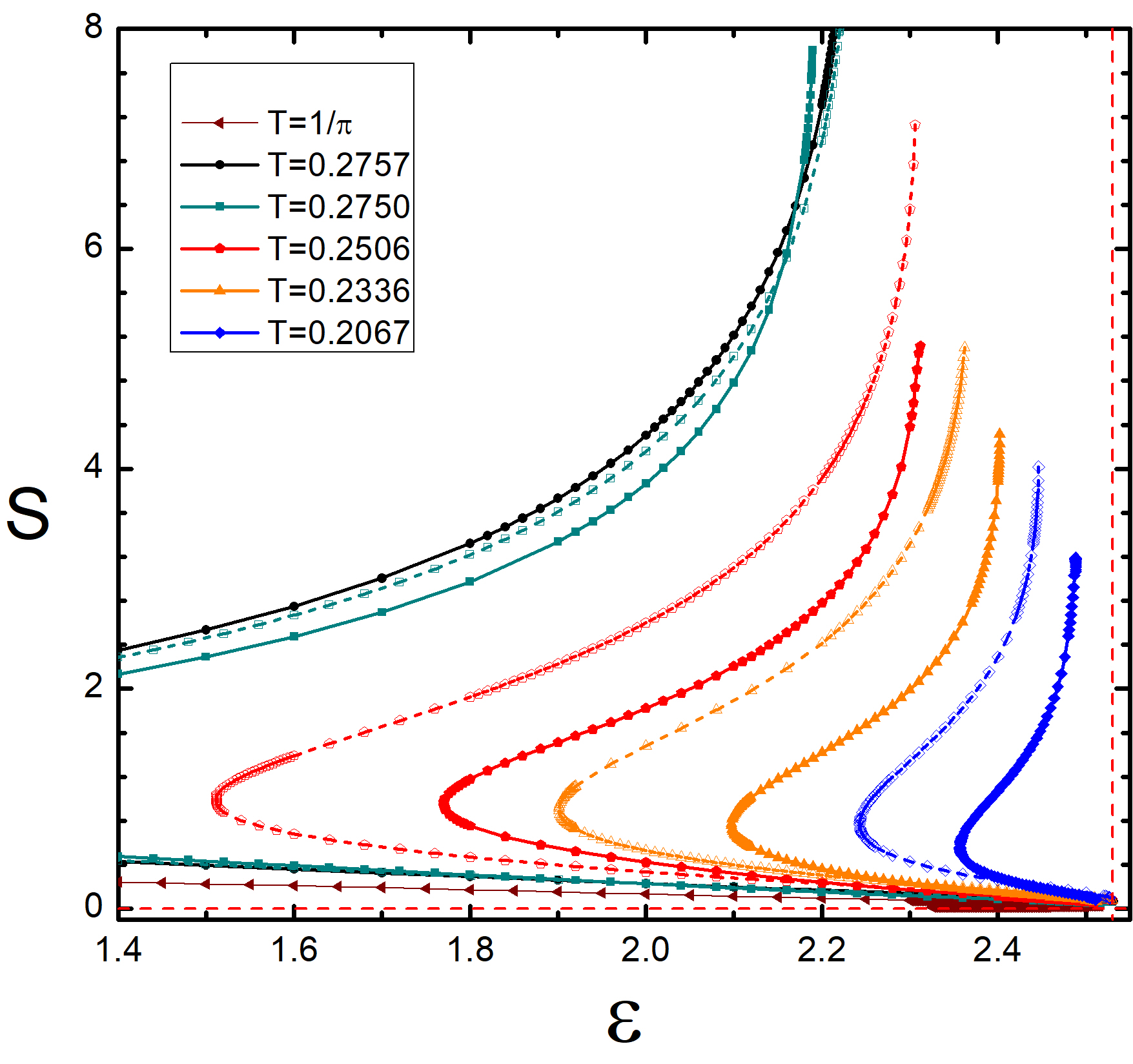}
\end{minipage}
\caption{\emph{Left}: Entropy against the boundary rotation parameter $\varepsilon$ for low temperature black holes with $T<T_{Schw}=\sqrt{3}/2\pi\simeq0.2757$. Two types of lines represent large (dashed line with hollow symbols) and small (solid line with solid symbols) black holes respectively, and the lines of same color denote the large and small black holes at the same temperature. \emph{Right}: The asymptotic behaviours for the curves of  entropy of large and small black holes. The black solid line represents the black holes with $T\simeq T_{Schw}$, and the brown solid line represents small black hole at $T=1/\pi$.   In both panels the vertical red dashed lines indicate $\varepsilon=\varepsilon_{c}$ and the horizontal red dashed lines indicate $S=0$.}
\label{fige13}
\end{figure}
For $T>T_{Schw}$, we first observe that the curves of entropy for two branches of black hole solutions at the fixed temperature are separate (The brown solid line represents the entropy of small black hole at $T=1/\pi$). When temperature is lowered to the $T_{Schw}\backsimeq0.2757$, the curves of entropy for two branches of black hole solutions combine into one curve (black solid line). As we continue to lower the temperature to $T<T_{Schw}$, we again obtain two separate curves of entropy at the fixed temperature, and each curve has two solutions with the fixed value of rotation parameter $\varepsilon$. The solid line with symbols denotes the small black holes, while the dashed line with hollow symbols denotes the large black holes.
\subsection{The horizon geometry}
\begin{figure}[t]
\centering
\includegraphics[scale=0.4]{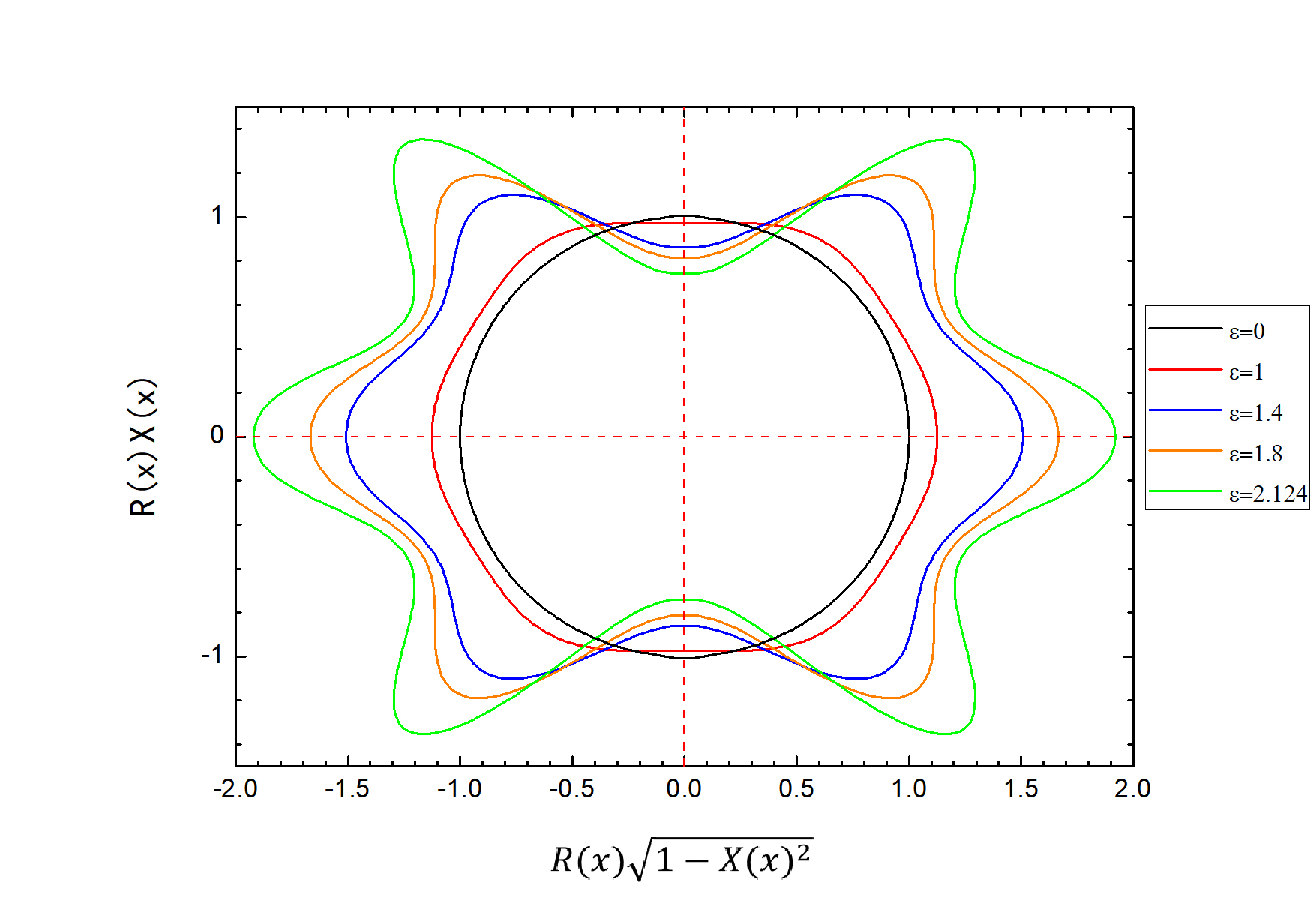}
\caption{Hyperbolic embedding of the cross sections of large black hole horizons for several values of the $\varepsilon$ at a fixed temperature $T=13/8\pi$. The black line represents represents the Schwarzschild-AdS$_{4}$ black hole.}
\label{fig7}
\end{figure}

\begin{figure}[t]
\centering
\begin{minipage}[c]{0.5\textwidth}
\centering
\includegraphics[scale=0.26]{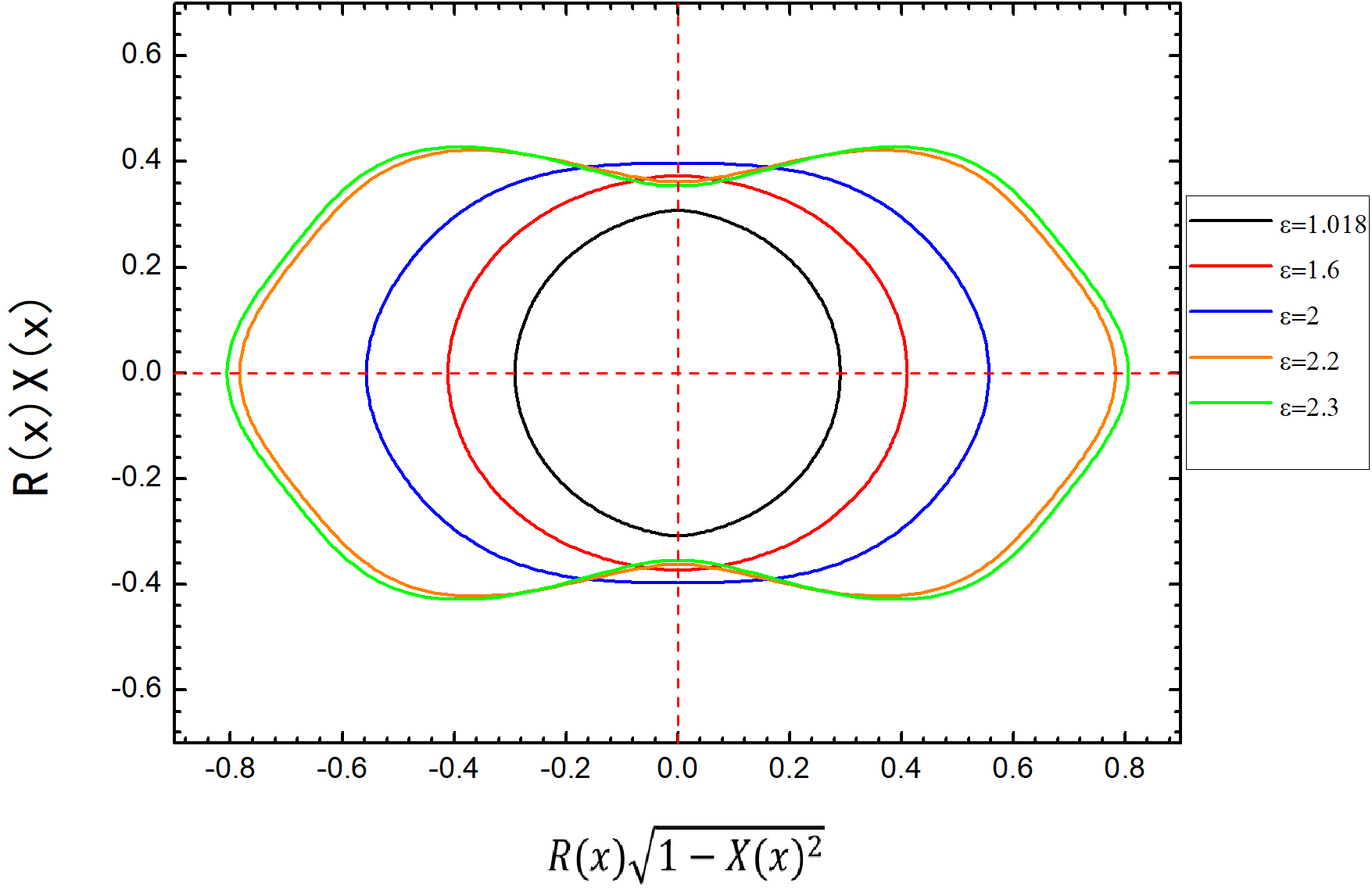}
\end{minipage}%
\begin{minipage}[c]{0.5\textwidth}
\centering
\includegraphics[scale=0.26]{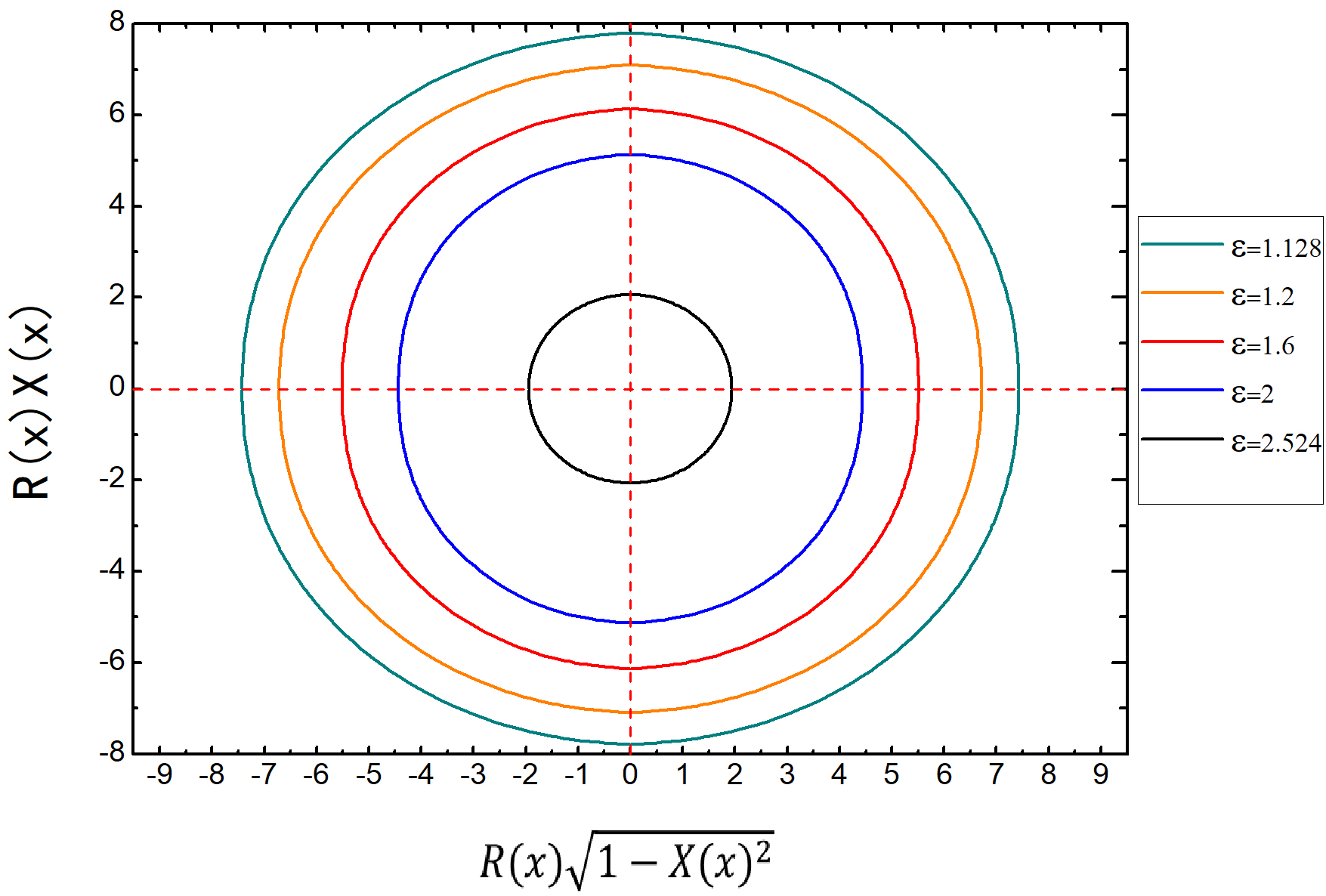}
\end{minipage}
\caption{Hyperbolic embedding of the cross sections of the black hole horizons for several values of $\varepsilon$ at the fixed temperature $T=0.2606$. \emph{Left}: The embedding for the large branch of black hole solutions with tripolar rotation. \emph{Right}: The embedding for the small branch of black hole solutions with tripolar rotation. }
\label{fig8}
\end{figure}

\hspace*{0.6cm}In order to obtain a better understanding of how the event horizon of deforming black hole behaves with increasing of the boundary rotation parameter $\varepsilon$, we could investigate the geometry of a two-dimensional surface in a curved space by using an isometric embedding in the three-dimensional space \cite{c1,c2,c3,c4,c5}, which has been introduced to study the horizon with dipolar differential rotation embedding in hyperbolic
space\cite{r1,r8}. In the polar coordinates, the metric of hyperbolic three-dimensional space $\mathbb{H}$$_{3}$ is given by
\begin{equation}\label{eq4.7}
ds_{\mathbb{H}^{3}}^{2}=\frac{dR^{2}}{1=R^{2}/\tilde{l}^{2}}+R^{2}\left[\frac{dX^{2}}{1-X^{2}}+(1-X^{2})d\phi^{2}\right],
\end{equation}
where $\tilde{l}$ is the radius of hyperbolic space. The induce metric on the horizon of the black hole with the metric (\ref{eq4.2}) is given by
\begin{equation}\label{eq4.8}
ds_{IM}^{2}=L^{2}\left[\frac{4y_{p}^{2}U_{3}(x,0)}{2-x^{2}}dx^{2}+y_{p}^{2}(1-x^{2})^{2}U_{5}(x,0)d\phi^{2}\right],
\end{equation}
and with the pull back of line element (\ref{eq4.7}) on the induced metric (\ref{eq4.8}), one could obtain a embedding of two-dimensional line element, which is given by a parameter curve $\{R(x),X(x)\}$
\begin{equation}\label{eq4.9}
ds_{RX}^{2}=\left[\frac{R'(x^{2})}{1+\frac{R(x)^{2}}{\tilde{l}^{2}}}+\frac{R(x)^{2}X'(x)^{2}}{1-X(x)^{2}}\right]dx^{2}+R(x)^{2}(1-X(x)^{2})d\phi^{2}.
\end{equation}
By using equation (\ref{eq4.8}) and (\ref{eq4.9}) we could obtain a first-order differential equation for the polar coordinate as follow
\begin{equation}
\begin{aligned}
4Q(x)D(x)(X(x)^{2}-1)^{2}\left[D(x)-\tilde{l}^{2}(X(x)^{2}-1)\right]
+4\tilde{l}^{2}D(x)X(x)(X(x)^{2}-1)D'(x)X'(x)\\
-(X(x)^{2}-1)^{2}\tilde{l}^{2}D'(x)^{2}-4P(x)^{2}(\tilde{l}^{2}+P(x))X'(x)^{2}=0,
\end{aligned}
\end{equation}
with $Q(x)=(2-x^{2})^{-1}(4y_{p}^{2}U_{3}(x,0))$ and $D(x)=y_{p}^{2}(1-x^{2})^{2}U_{5}(x,0)$. \\
\hspace*{0.6cm}We find it is hard to embed the horizon cross section into hyperbolic space $\mathbb{H}^{3}$ at the temperature $T=1/\pi$ with large values of  the rotation parameter $\varepsilon$ by using our numerical method, and the deformations of horizon are too small. So we recalculate the embedding of cross section and obtain the larger deformation of tripolar rotation at $T=13/8\pi$. In Fig.$\ $\ref{fig7}, we present the cross sections of the branch of large black hole horizons with tripolar rotation for several values of rotation parameter $\varepsilon$ at fixed temperature $T=13/8\pi$, and we set $\tilde{l}=0.73$ in all plots of this subsection. As we increase $\varepsilon$, the horizon cross section begins to deform and forms three hourglass shapes,  which could extend further at high values of $\varepsilon$.\\
\hspace*{0.6cm}In Fig.$\ $\ref{fig8}, we present the cross sections of horizons at a low temperature $T=0.2606$. In the left panel, we show the branch of large black hole solutions, and we find that the size of  the deformation of  horizon cross section with tripolar rotation
 decreases  as the temperature decreases. The right panel exhibits the branch of small black hole solutions, we could see that the cross sections of horizons become smaller with the increase of $\varepsilon$. Comparing with the curves of large black holes in the left panel, the curves of small black holes are nearly circular.
\subsection{The quasi-normal modes of black holes}
\begin{figure}[t]
\centering
\includegraphics[scale=0.3]{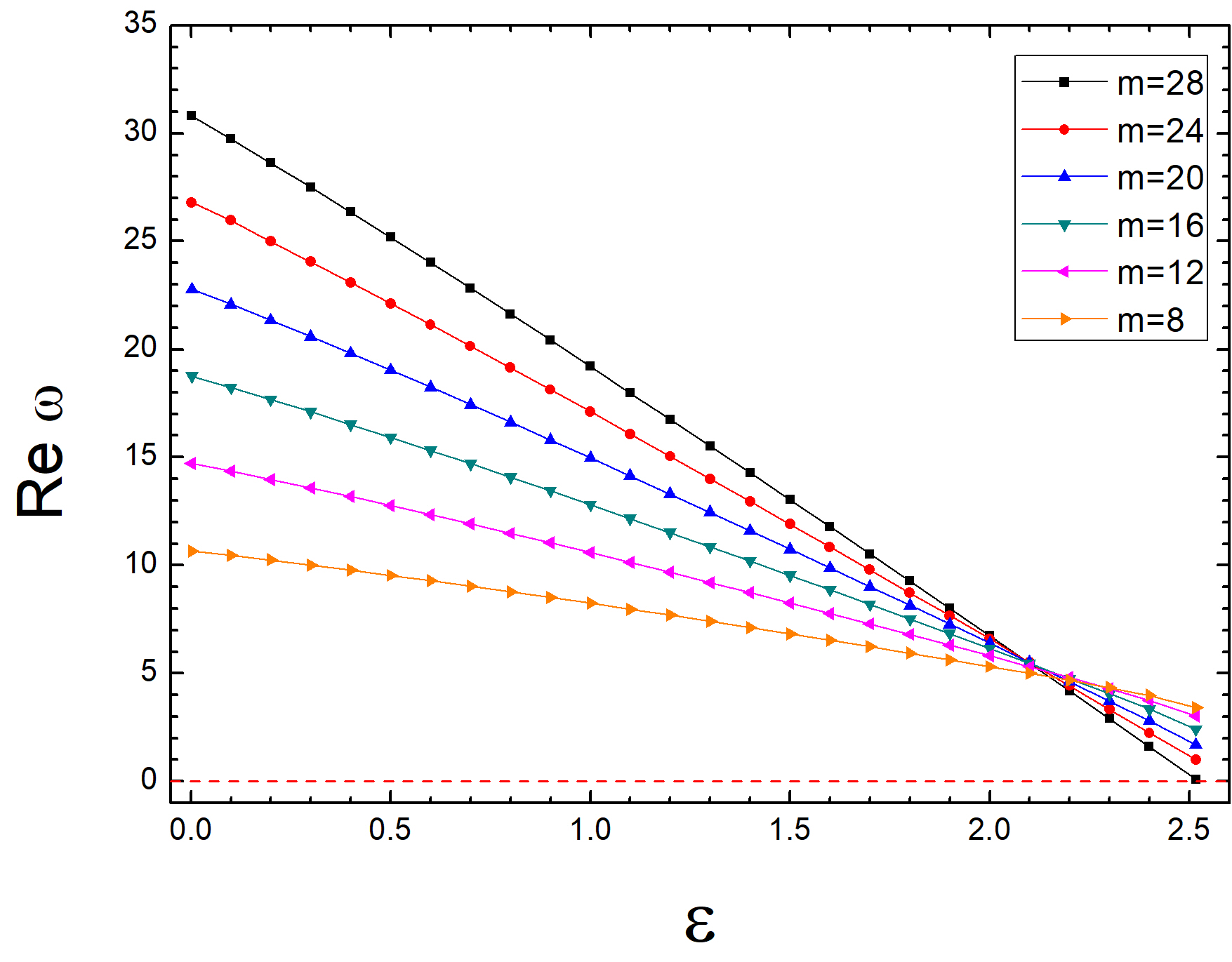}
\caption{The real part of frequencies $\omega$ against the rotation parameter $\varepsilon$ for small black hole at the temperature $T=1/\pi$. The red dashed line indicates Re $\,\omega=0 $.}
\label{fig6}
\end{figure}

\hspace*{0.6cm}In this subsection, we discuss  the linear stability of deforming black hole with tripolar differential rotation by studying the quasinormal modes. With the ansatz of the black hole metric (\ref{eq4.2}), the scalar field imposed regularity in ingoing Eddington-Finkelstein coordinates \cite{r7,c7} could be decomposed into
\begin{equation}
\Psi(t,x,y,\psi)=e^{i(m\psi-\omega t)}y^{-i\frac{2\omega y_{p}}{1+3y_{p}^{2}}}(1-y^{2})^{3}(1-x^{2})^{|m|}\xi(x,y),
\end{equation}
where the powers of $x$ and $y$ are chosen to make function $\xi(x,y)$ to be regular at the origin. The boundary conditions are imposed as follow: at $x=\pm1$ we require $\partial_{x}\xi(x,y)=0$, and at $y=0$, we require $\partial_{y}\xi(x,y)=0$. At $y=1$, we require
\begin{equation}
-2i\,y_{p}\,\omega\, \xi(x,y)+(1+3y_{p}^{2})\partial_{y}\xi(x,y)=0.
\end{equation}
\hspace*{0.6cm}In Fig.$\ $\ref{fig6}, for a small black hole, we plot the real part of the quasi-normal frequencies $\omega$ as the function of rotation  parameter $\varepsilon$ for the corresponding values of $m$ at the fixed temperature $T=1/\pi$, and the red dashed line indicates Re $\omega=0$.
The frequencies Re $\omega$ with $m\leqslant28$ are always positive values in the spectrum of perturbations, and the frequency begins to be negative at a specific value of $\varepsilon$ when $m\geqslant29$. The characteristic of Re $\omega$ against the boundary rotation parameter $\varepsilon$ is similar to that of soliton solutions. For the first stable mode at $m=29$, one can expect some branches of black hole with scalar hair $\Psi$ condensation could be found.
\section{Conclusions}\label{sec5}
\hspace*{0.6cm}In this paper, we analyzed the conformal boundary of four dimensional static asymptotically AdS solutions in Einstein gravity and constructed the numerical solutions of solitons and black holes with odd multipolar differential rotation boundary. Comparing with the dipolar differential rotation solutions in \cite{r1}, we found that the norm of Killing vector $\partial_{t}$ becomes spacelike for the certain regions of $\theta$ with $\varepsilon\in(2,2.124)$, solitons and black holes with tripolar differential rotation do not develop hair due to superradiance at high temperature with $T>T_{Schw}\simeq0.2757$, which was different from the case of dipolar rotation. For the large black holes of the high temperature, we found that the maximum value of $\varepsilon$  decreases with the increases of temperature. For $T\in(1.5732,2/\pi)$, the maximum values of $\varepsilon$ are smaller than $2$,
meanwhile, the Killing vector $\partial_{t}$ is timelike  for some regions of $\varepsilon<2$, solitons and black holes with tripolar differential rotation will develop hair due to superradiance. When temperature was lowered, we found that the entropy for large and small branches of black holes solutions firstly combine into one curve at $T=T_{schw}\simeq0.2757$ and then separate into two curves for $T<T_{schw}$, the entropy for large and small branches of black hole solutions have two solutions at a fixed value of $\varepsilon$. These were different from the results in the cases of dipolar and quadrupolar differential rotation boundaries \cite{r1,Li}. Furthermore, with the isometric embedding of horizon, it is clearly seen that black hole horizon is deformed into three hourglass shapes.  By studying the quasinormal modes, we discussed the linear stability of deforming solitons and black holes with tripolar rotation, respectively, and found that for some branches of solution with scalar hair $\Psi$ condensation, the minimal azimuthal harmonic index $m$ is equal to 24 for solitons and $m=29$ for black holes.\\
\hspace*{0.6cm}At fixed temperature $T=1/\pi$, it is interesting to find that though the norm of Killing vector $\partial_{t}$ becomes spacelike for certain regions of $\theta$ with $\varepsilon>2$, solitons and black holes with tripolar differential rotation still exist and do not develop hair due to superradiance. When temperature is much higher than $T_{schw}$, we also found that even though the norm of Killing vector $\partial_{t}$ keeps timelike for some regions of $\varepsilon<2$, solitons and black holes with tripolar differential rotation could
be unstable and develop hair due to superradiance, which is different from the cases of solutions with dipolar and quadrupolar differential rotation boundaries in \cite{Li,r1}.  We checked the numerical solutions with pentapolar differential rotation, which shows very similar results to tripolar differential rotation. There exists the solution of pentapolar differential rotation when $\varepsilon>2$. \\
\hspace*{0.6cm}At present, we have studied the behaviours of solitons and black holes with tripolar differential rotation, but the angular momentum, energy densities and thermodynamic properties of black holes with odd multipolar differential rotation have not been studied,  and  we hope to investigate these in our future work. Besides, we have two extensions of our work. The one extension of our work is to study the action of Einstein-Maxwell gravity in AdS spacetime and construct the deforming charged black holes. We find there are three branches of solutions because of the existence of charges. The phase diagram of solutions is more complicated than that without charges. The another is to extend the deforming black holes to $f(R)$ gravity and Gauss-Bonnet theory.

\section*{Acknowledgement}
\hspace*{0.6cm}We would like to thank  Yu-Xiao Liu, Jie Yang and Li Zhao  for  helpful discussion.
Some computations were performed on the   shared memory system at  institute of computational physics and complex systems in Lanzhou university. This work was supported by the Fundamental Research Funds for the Central Universities (Grants No. lzujbky-2017-182).

\end{document}